\documentclass[preprint,showpacs,superscriptaddress]{revtex4}
\usepackage[pdftex]{graphicx}
\usepackage{amssymb,amsfonts,amsmath,verbatim}
\usepackage{epstopdf}
\usepackage{setspace}

\begin{document}
\title{Negative differential conductance and
super-Poissonian shot noise in single-molecule magnet junctions}

\author{Hai-Bin Xue}
\email{xuehaibin@tyut.edu.cn}
\affiliation{College of Physics and Optoelectronics, Taiyuan University of Technology,
Taiyuan 030024, China}
\author{Jiu-Qing Liang}
\affiliation{Institute of Theoretical Physics, Shanxi University, Taiyuan 030006,
China}
\author{Wu-Ming Liu}
\email{wliu@iphy.ac.cn}
\affiliation{Beijing National Laboratory for Condensed Matter Physics, Institute of
Physics, Chinese Academy of Sciences, Beijing 100190, China}

\date{\today}

\begin{abstract}
Molecular spintroinic device based on a single-molecule magnet is one of the ultimate goals of semiconductor nanofabrication technologies.
It is thus necessary to understand the electron transport properties of a single-molecule magnet junction.
Here we study the negative differential conductance and
super-Poissonian shot noise properties of electron transport through a single-molecule magnet weakly coupled
to two electrodes with either one or both of them being ferromagnetic. We predict that the negative differential conductance
and super-Poissonian shot noise, which can be tuned by a gate voltage,
depend sensitively on the spin polarization of the source and drain
electrodes. In particular, the shot noise in the negative differential conductance region can be enhanced
or decreased originating from the different formation mechanisms of
negative differential conductance. The effective competition between fast and slow transport channels is
responsible for the observed negative differential conductance and super-Poissonian shot noise. In addition,
we further discuss the skewness and kurtosis properties of transport current in the
super-Poissonian shot noise regions. Our findings suggest a tunable negative differential conductance molecular device,
and the predicted properties of high-order current cumulants are very interesting for a better understanding of
electron transport through single-molecule magnet junctions.
\end{abstract}
\keywords{single-molecule magnet; negative differential conductance;
super-Poissonian shot noise}
\pacs{75.50.Xx, 72.70.+m, 73.63.-b}

\maketitle


\section*{Introduction}

Electronic transport through a single-molecule magnet (SMM) has been
intensively studied both experimentally\cite%
{Heersche,Jo,Grose,Loth,Zyazin,Roch,Komeda,Kahle,Vincent,Thiele} and
theoretically\cite%
{Romeike01,Romeike02,Timm01,Elste01,Elste02,Timm02,Wegewijs,Gonzalez01,Gonzalez02,Misiorny02,Misiorny03,Xiejap12,Misiorny04,Xiejap13,WangRN}
due to its applications in molecular spintronics\cite{Lapo}, but these
investigations were focused mainly on the differential conductance or
average current. Although the shot noise of electron transport
through a SMM has not yet been observed experimentally, new techniques
based on carbon nanotubes have been proposed for its possible realization%
\cite{Gruneis}. Recently, the current noise properties of electron transport
through a SMM have been attracting much theoretical
research interests\cite%
{Romeike03,Imura,Misiorny01,Xuejap10,Xuepla,Xuejap11,WangRQ,Aguado} due to they can
provide a deeper insight into the nature of transport mechanisms that cannot be
obtained by measuring the differential conductance or average current\cite%
{Blanter,Nazarov}. For example, the super-Poissonian shot noise can be used
to reveal the information about the internal level structure of the SMM, the
left-right asymmetry of the SMM-electrode coupling\cite{Xuejap10,Xuepla},
and the angle between the applied magnetic field and the SMM's easy axis\cite%
{Xuejap11}; and distinguish the two types of different nonequilibrium
dynamics mechanisms, namely, the quantum tunneling of magnetization process
and the thermally excited spin relaxation\cite{WangRQ}. In particular, the
frequency-resolved shot noise spectrum of artificial SMM, e.g., a CdTe
quantum dot doped with a single $S=5/2$ Mn spin, can allow one to separately
extract the hole and Mn spin relaxation times via the Dicke effect\cite%
{Aguado}.

Among these observed or predicted characteristics, the negative differential
conductance (NDC) is especially concerned due to the SMM's potential applications in a new generation of
molecule-based memory devices and logic circuits. On the other hand, the shot noise is usually
the sub-Poissonian statistics in non-interacting fermion systems originating from the Pauli exclusion principle.
Thus, the super-Poissonian shot noise is another important characteristic of transport current. For
the SMM weakly coupled to two normal metal electrodes, the NDC formation mechanism originates essentially
from the non-equilibrium electron occupation of the system eigenstates
entering bias voltage window\cite{Timm01,Xuejap10}, namely, the
increased current magnitudes of the new opened transport channels do not
compensate the decreased current magnitude(s) of the already opened transport
channel(s), and the shot noise in this NDC region is obviously enhanced
even up to a super-Poissonian shot noise value. In particular, the occurrence of
super-Poissonian shot noise depends on the effective competition
between different transport channels, thus, the SMM's internal level
structure and the left-right asymmetry of the SMM-electrode coupling, which
can tune the SMM transport channels, have an important influence on the
super-Poissonian shot noise properties\cite{Xuejap10,Xuepla,Xuejap11}.
Whereas for the SMM weakly coupled to two electrodes with either one or both
of them being ferromagnetic, the spin polarization of the source and drain
electrodes play an important role in the forming speed of the correlated
SMM eigenstates involved in the electron tunneling processes, and thus have a
remarkable influence on the transport channels entering bias voltage window%
\cite{Misiorny01,Misiorny02,LuoW}. Consequently, the spin polarization of
the source and drain electrodes will have an significant impact on the NDC
and super-Poissonian shot noise properties of this SMM system. However, the
influences of the spin polarization of the source and drain electrodes on the
NDC and super-Poissonian shot noise in the SMM system have not yet been
revealed.

The goal of this report is thus to study the influences of the spin
polarization of the source and drain electrodes and the applied gate voltage
on the NDC and super-Poissonian shot noise in a SMM weakly coupled to two
electrodes with either one or both of them being ferromagnetic, and discuss
the underlying mechanisms of the observed NDC and super-Poissonian shot
noise. It is demonstrated that the gate-voltage-controlled NDC and
super-Poissonian shot noise depend sensitively on the spin polarization of
the source and drain electrodes. In particular, whether the shot noise in the NDC
region being enhanced or not is associated with the formation mechanism of
the NDC. Moreover, the skewness and kurtosis in the super-Poissonian shot noise regions
show the crossovers from a large positive (negative) to a large negative (positive)
values, which also depend on the spin polarization of the source and drain electrodes.
These observed characteristics are very interesting for a better understanding of
electron transport through single-molecule magnet junctions and will allow for
experimental tests in the near future.

\section*{Results}
\subsection*{Single-molecule magnet junction}
The SMM junction consists of a SMM weakly coupled to two electrodes, see Fig. 1.
The SMM is characterized by the lowest unoccupied non-degenerate molecular orbital
(LUMO), the phenomenological giant spin $\overrightarrow{S}$, and the uniaxial anisotropy.
The SMM Hamiltonian is thus described by
\begin{equation}
H_{\text{SMM}}=(\varepsilon _{d}-eV_{g})\hat{n}+\frac{U}{2}\hat{n}(\hat{n}%
-1)-J\,\vec{s}\cdot \vec{S}-K_{2}(S_{z})^{2}-B_{z}(s^{z}+S^{z}),  \label{model}
\end{equation}%
Here, the first two terms depict the LUMO, $\hat{n}\equiv d_{\uparrow
}^{\dag }d_{\uparrow }+d_{\downarrow }^{\dag }d_{\downarrow }$ and $U$ are
respectively the electron number operator and the Coulomb repulsion between
two electrons in the LUMO, with $d_{\sigma }^{\dag }$ ($d_{\sigma }$) being
the electron creation (annihilation) operators with spin $\sigma $ and
energy $\varepsilon _{d}$ (which can be tuned by a gate voltage $V_{g}$) in
the LUMO. The third term describes the exchange coupling between the
conduction electron spin $\vec{s}\equiv \sum_{\sigma \sigma ^{\prime
}}d_{\sigma }^{\dag }\left( \vec{\sigma}_{\sigma \sigma ^{\prime }}\right)
d_{\sigma ^{\prime }}$ in the LUMO and the SMM spin $\vec{S}$, with $\vec{%
\sigma}\equiv $ $(\sigma _{x},\sigma _{y},\sigma _{z})$ being the vector of
Pauli matrices. The forth term stands for the anisotropy energy of the SMM
whose easy-axis is $Z$-axis ($K_{2}>0$). The last term denotes Zeeman
splitting. For simplicity, we assume an external magnetic field $%
\overrightarrow{B}$ is applied along the easy axis of the SMM.

The relaxation in the two electrodes is assumed to be sufficiently fast so
that their electron distributions can be described by equilibrium Fermi
functions. The two electrodes are thus modeled as noninteracting Fermi gases
and the corresponding Hamiltonians read
\begin{equation}
H_{\text{Leads}}=\sum_{\alpha \mathbf{k}s}\varepsilon _{\alpha \mathbf{k}%
s}a_{\alpha \mathbf{k}s}^{\dag }a_{\alpha \mathbf{k}s},  \label{Leads}
\end{equation}%
where $a_{\alpha \mathbf{k}\sigma }^{\dag }$ ($a_{\alpha \mathbf{k}\sigma }$%
) is the electron creation (annihilation) operators with energy $%
\varepsilon _{\alpha \mathbf{k}\sigma }$, momentum $\mathbf{k}$ and spin $s$
in $\alpha $ ($\alpha =L,R$) electrode, and the index $s=+\left( -\right) $
denotes the majority (minority) spin states with the density of states $%
g_{\alpha }^{s}$. The electrode polarization is characterized by the orientation
of the polarization vector $\mathbf{p}_{\alpha }$ and its magnitude is
defined as $p_{\alpha }=(g_{\alpha }^{\uparrow }-g_{\alpha }^{\downarrow
})/(g_{\alpha }^{\uparrow }+g_{\alpha }^{\downarrow })$. Here, the
polarization vectors $\mathbf{p}_{L}$ (left electrode) and $\mathbf{p}_{R}$
(right electrode) are parallel to the spin quantization $Z$ axis, and
spin-up $\uparrow $ and spin-down $\downarrow $ are respectively defined to
be the majority spin and minority spin of the ferromagnet. The tunneling
between the SMM and the two electrodes are thus described by
\begin{equation}
H_{\text{tun}}=\sum_{\alpha \mathbf{k}\sigma }\left( t_{\alpha \mathbf{k}%
\sigma }a_{\alpha \mathbf{k}\sigma }^{\dag }d_{\sigma }+\text{H.c.}\right) ,
\label{tunneling}
\end{equation}%
Here, for the ferromagnetic electrode case, the electronic tunneling rates depend
on the conduction-electron spin, namely, $\Gamma _{\alpha }^{\uparrow }=2\pi
|t_{\alpha }|^{2}g_{\alpha }^{\uparrow }=(1+p_{\alpha })\Gamma _{\alpha }/2$
and $\Gamma _{\alpha }^{\downarrow }=2\pi |t_{\alpha }|^{2}g_{\alpha
}^{\downarrow }=(1-p_{\alpha })\Gamma _{\alpha }/2$, where the tunneling
amplitudes $t_{\alpha }$ and the density of the state $g_{\alpha }^{\sigma }$
are assumed to be independent of wave vector and energy, and $\Gamma
_{\alpha }=\Gamma _{\alpha }^{\uparrow }+\Gamma _{\alpha }^{\downarrow }$;
while for the normal-metal electrode case, $p_{\alpha }=0$, thus, $\Gamma
_{\alpha }^{\uparrow }=\Gamma _{\alpha }^{\downarrow }=$ $\Gamma _{\alpha }/2
$.

In addition, we assume that the bias voltage is symmetrically entirely
dropped at the SMM-electrode tunnel junctions, i.e., $\mu _{L}=-\mu
_{R}=V_{b}/2$, which implies that the levels of the SMM are independent of
the applied bias voltage, and choose meV as the unit of energy. In the Coulomb blockade regime, the occurrence or
absence of super-Poissonian shot noise is related to the sequential
tunneling gap $\epsilon _{se}$ that being the energy difference between the
ground state of charge $N$ and the first excited state of charge $N-1$, and
the \textquotedblleft vertical\textquotedblright\ energy gap $\epsilon
_{co} $ between the ground state of charge $N$ and the first excited state
of the same charge\cite{Aghassi01}. In the present work, we only study the
electron transport above the sequential tunneling threshold, namely, $%
V_{b}>2\epsilon _{se}$. In this bias voltage region, the conduction
electrons have sufficient energy to overcome the Coulomb blockade and tunnel
sequentially through the SMM. It should be noted that the transport current in
the Coulomb blockade regime is exponentially suppressed and the co-tunneling
tunneling process is dominant in the electron transport, thus, the normalized shot
noise will deviate from the present results when taking co-tunneling into
account. The parameters of the SMM are chosen as: $S=2$, $\varepsilon
_{d}=0.2$, $U=0.1$, $J=0.1$, $K_{2}=0.04$, $B=0.08$, $\Gamma _{L}=\Gamma
_{R}=\Gamma=0.002$ and $k_{B}T=0.02$.

We first study numerically the effects of the spin polarization
of the two electrodes and the applied gate voltage on the NDC and
super-Poissonian shot noise in the three different electrode configurations (see Fig. 1), namely,
(i) the ferromagnetic lead (Source) - SMM - normal-metal lead (Drain) (i.e., the F-SMM-N system),
(ii) the normal-metal lead (Source) - SMM - ferromagnetic lead (Drain) (i.e., the N-SMM-F system),
(iii) the ferromagnetic lead (Source) - SMM - ferromagnetic lead (Drain) (i.e., the F-SMM-F system).

\subsection*{The ferromagnetic lead (Source) - SMM - normal-metal lead (Drain)}

For the F-SMM-N system considered here, the conduction
electron will tunnel into the SMM from the ferromagnetic lead and then
tunnel out of the SMM onto the normal-metal lead. The strengths of
tunneling coupling of the SMM with two electrodes can be expressed as $%
\Gamma _{L}^{\uparrow }=\Gamma (1+p_{L})/2$, $\Gamma _{L}^{\downarrow
}=\Gamma (1-p_{L})/2$ and $\Gamma _{R}^{\uparrow }=\Gamma _{R}^{\downarrow
}=\Gamma /2$. Since only the energy eigenvalues of singly-occupied and
doubly-occupied eigenstates $\epsilon ^{\pm }\left( 1,m\right) $ and $%
\epsilon \left( 2,m\right) $ depend on the gate voltage $V_{g}$, the
transition between the singly- and doubly-occupied eigenstates, or between the
empty- and singly-occupied eigenstates first entering bias voltage
window can be tuned by the gate voltage\cite{Timm01}. For example, for a relatively
small or negative gate voltage, the transition from the singly- to
empty-occupied eigenstates first takes place; while for a large enough gate
voltage that from the double- to singly-occupied eigenstates first occurs.

Figures 2(a) and 2(b), 2(e) and 2(f) show the average current and shot noise as a function of the bias
voltage for different gate voltages $V_{g}$ with $p_{L}=0.3$ and $p_{L}=0.9$.
For a large enough spin polarization of source electrode $p_{L}$, the
super-Poissonian shot noise is observed when the transition from the doubly-
and singly-occupied eigenstates first participates in the electron transport
with the bias voltage increasing, see the short dashed, short dash-dotted and
thick dashed lines in Fig. 2(f), whereas for the QD system the
super-Poissonian noise dose not appear\cite{Lindebaum}. This characteristic
can be understood in terms of the effective competition between fast and
slow transport channels\cite%
{Xuejap10,Xuepla,Xuejap11,Aghassi01,Safonov,Djuric,Aghassi02,Xueepjb,Xueaip,Xueaop}
and the forming speed of the new correlated eigenstates\cite{Xueaip}. The current
magnitudes of the SMM transport channels can be expressed as\cite%
{Timm01,Xuejap10,Xuejap11}%
\begin{equation}
I_{\left\vert n,m\right\rangle \longrightarrow \left\vert
n-1,m-1/2\right\rangle }=C_{\left\vert n-1,m-1/2\right\rangle ,\left\vert
n,m\right\rangle }\Gamma _{R}^{\uparrow }n_{R}^{\left( -\right) }\left(
\epsilon _{\left\vert n,m\right\rangle }-\epsilon _{\left\vert
n-1,m-1/2\right\rangle }-\mu _{R}\right) P_{\left\vert n,m\right\rangle },
\label{channelup}
\end{equation}%
\begin{equation}
I_{\left\vert n,m\right\rangle \longrightarrow \left\vert
n-1,m+1/2\right\rangle }=C_{\left\vert n-1,m+1/2\right\rangle ,\left\vert
n,m\right\rangle }\Gamma _{R}^{\downarrow }n_{R}^{\left( -\right) }\left(
\epsilon _{\left\vert n,m\right\rangle }-\epsilon _{\left\vert
n-1,m+1/2\right\rangle }-\mu _{R}\right) P_{\left\vert n,m\right\rangle },
\label{channeldown}
\end{equation}%
where $C_{\left\vert n-1,m\pm 1/2\right\rangle ,\left\vert n,m\right\rangle
}=\left\vert \left\langle n-1,m\pm 1/2\right\vert d_{\sigma }\left\vert
n,m\right\rangle \right\vert ^{2}$ is a constant which related to the two
SMM eigenstates but independent of the applied bias voltage, and $%
P_{\left\vert n,m\right\rangle }$ is the occupation probability of the SMM
eigenstate $\left\vert n,m\right\rangle $. Here, the Fermi function $%
n_{R}^{\left( +\right) }\left( \epsilon _{\left\vert n,m\right\rangle
}-\epsilon _{\left\vert n-1,m-1/2\right\rangle }-\mu _{R}\right) $ changes
very slowly with increasing bias voltage above the sequential tunneling
threshold, namely, $n_{R}^{\left( +\right) }\left( \epsilon _{\left\vert
n,m\right\rangle }-\epsilon _{\left\vert n-1,m-1/2\right\rangle }-\mu
_{R}\right) \simeq 0$, thus, $n_{R}^{\left( -\right) }\left( \epsilon
_{\left\vert n,m\right\rangle }-\epsilon _{\left\vert n-1,m-1/2\right\rangle
}-\mu _{R}\right) \simeq 1$. The current magnitude of the SMM transport channel is thus mainly
determined by the occupation probability $P_{\left\vert n,m\right\rangle }$
and $\Gamma _{R}^{\sigma }$.

In order to give a qualitative explanation for the underlying mechanism of
the observed super-Poissonian shot noise, we plot the occupation probabilities
of the SMM eigenstates as a function of bias voltage for $p_{L}=0.9$ and $%
V_{g}=0.6$ in Fig. 3. With increasing bias voltage, the transport channel $%
\left\vert 2,2\right\rangle \overset{\downarrow }{\longrightarrow }%
\left\vert 1,5/2\right\rangle $ begins to participate in the electron
transport. When the bias voltage increases up to about 0.6 meV, the new transport
channel $\left\vert 2,2\right\rangle \overset{\uparrow }{\longrightarrow }%
\left\vert 1,3/2\right\rangle ^{-}$ enters the bias voltage window. In this
situation, the conduction electron can tunnel out SMM via the two transport
channels $\left\vert 2,2\right\rangle \overset{\downarrow }{\longrightarrow }%
\left\vert 1,5/2\right\rangle $ and $\left\vert 2,2\right\rangle \overset{%
\uparrow }{\longrightarrow }\left\vert 1,3/2\right\rangle ^{-}$. For the
F-SMM-N system, the electron tunneling between the SMM and the drain
electrode (normal-metal lead) is independent of the conduction electron
spin, thus the tunneling process mainly relies on the forming speed of the
new doubly-occupied eigenstate $\left\vert 2,2\right\rangle $. In the case
of $\Gamma _{L}^{\uparrow }\gg \Gamma _{L}^{\downarrow }$, a new
doubly-occupied eigenstate $\left\vert 2,2\right\rangle $\ can be quickly
formed when the spin-up electron tunnels out of the SMM; whereas for the
case of the spin-down electron tunneling out of the SMM,\ the forming of the
corresponding new doubly-occupied eigenstate $\left\vert 2,2\right\rangle $
takes a relatively longer time. Thus, for a large enough $p_{L}$, the
fast transport channel $\left\vert 2,2\right\rangle \overset{\uparrow }{%
\longrightarrow }\left\vert 1,3/2\right\rangle ^{-}$ can be modulated by the
correlated slow channel $\left\vert 2,2\right\rangle \overset{\downarrow }{%
\longrightarrow }\left\vert 1,5/2\right\rangle $, which leads to the
bunching effect of the conduction electrons being formed, and is responsible
for the formation of the super-Poissonian noise. When $V_{b}>0.9$, the
transport channels $\left\vert 1,5/2\right\rangle \overset{\uparrow }{%
\longrightarrow }\left\vert 0,2\right\rangle $ and $\left\vert
1,3/2\right\rangle ^{-}\overset{\downarrow }{\longrightarrow }\left\vert
0,2\right\rangle $ enter the bias volatge window, so that the two successive
electron tunneling processes $\left\vert 2,2\right\rangle \overset{%
\downarrow }{\longrightarrow }\left\vert 1,5/2\right\rangle \overset{%
\uparrow }{\longrightarrow }\left\vert 0,2\right\rangle $ and $\left\vert
2,2\right\rangle \overset{\uparrow }{\longrightarrow }\left\vert
1,3/2\right\rangle ^{-}\overset{\downarrow }{\longrightarrow }\left\vert
0,2\right\rangle $ can be formed. Consequently, the formed active
competition between the fast-and-slow transport channels is suppressed even
destroyed with the current magnitudes of the two new transport channels
increasing, which leads to the super-Poissonian shot noise being decreased
and even to the sub-Poissonian.

\subsection*{The normal-metal lead (Source) - SMM - ferromagnetic lead (Drain)}

In the N-SMM-F system, the strengths of tunnel coupling between the SMM and
the two electrodes are described by $\Gamma _{L}^{\uparrow }=\Gamma
_{L}^{\downarrow }=\Gamma /2$, $\Gamma _{R}^{\uparrow }=\Gamma (1+p_{R})/2$,
$\Gamma _{R}^{\downarrow }=\Gamma (1-p_{R})/2$. It is demonstrated that the
NDC is observed for a small enough or negative gate voltage, and a relatively
large spin polarization of drain electrode $p_{R}$, see the solid and dashed
lines in Figs. 4(a) and 4(e), especially for a large enough spin
polarization $p_{R}$ a strong NDC takes place, see the solid and dashed
lines in Fig. 4(e). Moreover, the shot noise can be significantly enhanced
and reaches up to a super-Poissonian value when the magnitude of the total
current begins to decrease, but the super-Poissonian value in the NDC region
is then decreased with further increasing the bias voltage, see the solid
and dashed lines in Figs. 4(b) and 4(f). While for a large enough gate
voltage, the peaks of super-Poissonian shot noise are observed for a
relatively large spin polarization $p_{R}$, see the short
dash-dotted and thick dashed lines in Figs. 4(b) and 4(f). The observed NDC and
super-Poissonian shot noise characteristics can also be attributed to the
mechanism of the fast-and-slow transport channels. Here, we take the $%
V_{g}=-0.1$ and $V_{g}=0.6$ cases with $p_{R}=0.9$ as examples to illustrate
these characteristics.

For the $V_{g}=-0.1$ case, the transition from singly-occupied to empty
eigenstates $\left\vert 1,5/2\right\rangle \overset{\uparrow }{%
\longrightarrow }\left\vert 0,2\right\rangle $ first participates in the
electron transport with increasing the bias voltage, see Figs. 5(a) and 5(b). When the
bias voltage is larger than 0.33 meV, the SMM has a small probability of
forming the empty-occupied eigenstate $\left\vert 0,-2\right\rangle $, see
the thick solid line in Fig. 5(a). If the spin-down electron tunnels into
the SMM, the singly-occupied eigenstate $\left\vert 1,-5/2\right\rangle $
can be formed. In this case, for a large enough spin polarization $p_{R}$,
namely, $\Gamma _{R}^{\uparrow }\gg \Gamma _{R}^{\downarrow }$, the spin-down
electron will remain for a relatively long time in the SMM, so that the
electron tunneling processes via the fast transport channels $\left\vert
1,5/2\right\rangle \overset{\uparrow }{\longrightarrow }\left\vert
0,2\right\rangle $ and $\left\vert 1,-3/2\right\rangle ^{-}\overset{\uparrow
}{\longrightarrow }\left\vert 0,-2\right\rangle $ can be blocked and the
conduction electrons appear the bunching effect. On the other hand, the
current magnitude of the formed fast transport channel $\left\vert
1,5/2\right\rangle \overset{\uparrow }{\longrightarrow }\left\vert
0,2\right\rangle $ begins to decrease with increasing the bias voltage up to
about 4.25 meV, while that of the two new opened transport channels $%
\left\vert 1,-3/2\right\rangle ^{-}\overset{\uparrow }{\longrightarrow }%
\left\vert 0,-2\right\rangle $ and $\left\vert 1,-5/2\right\rangle \overset{%
\downarrow }{\longrightarrow }\left\vert 0,-2\right\rangle $ increase. Since
the occupation probabilities of the eigenstates $\left\vert
1,-3/2\right\rangle ^{-}$ and $\left\vert 1,-5/2\right\rangle $, $%
P_{\left\vert 1,-3/2\right\rangle ^{-}}$ and $P_{\left\vert
1,-5/2\right\rangle }$ are much smaller that $P_{\left\vert
1,5/2\right\rangle }$, see Fig. 5(b), thus, for the $\Gamma _{R}^{\uparrow
}\gg \Gamma _{R}^{\downarrow }$ case the decreased current magnitude of
transport channel $\left\vert 1,5/2\right\rangle \overset{\uparrow }{%
\longrightarrow }\left\vert 0,2\right\rangle $ is much larger than the
increased current magnitudes of transport channels $\left\vert
1,-3/2\right\rangle ^{-}\overset{\uparrow }{\longrightarrow }\left\vert
0,-2\right\rangle $ and $\left\vert 1,-5/2\right\rangle \overset{\downarrow }%
{\longrightarrow }\left\vert 0,-2\right\rangle $. Thus, a strong NDC is
observed, see the solid line in Fig. 4(e). Moreover, the active competition
between the fast channel of current decreasing and the slow channels of
current increasing can also obviously enhance the shot noise. Consequently, the shot
noise is significantly enhanced by the above two mechanisms and reaches up
to a very large value of super-Poissonian shot noise before the occupation
probabilities $P_{\left\vert 1,-3/2\right\rangle ^{-}}$ and $P_{\left\vert
1,-5/2\right\rangle }$ are larger than $P_{\left\vert 1,5/2\right\rangle }$,
and $P_{\left\vert 0,-2\right\rangle }$ is larger than $P_{\left\vert
0,2\right\rangle }$. With the bias voltage further increasing, the value of super-Poissonian shot
noise is decreased quickly but still remains the super-Poissonian
distribution. This originates from the fact that the transport channels $%
\left\vert 1,-3/2\right\rangle ^{-}\overset{\uparrow }{\longrightarrow }%
\left\vert 0,-2\right\rangle $ and $\left\vert 1,-5/2\right\rangle \overset{%
\downarrow }{\longrightarrow }\left\vert 0,-2\right\rangle $ can form a
new effective competition between the fast and slow transport channels. When the occupation
probability $P_{\left\vert 2,2\right\rangle }$ is larger than $P_{\left\vert
0,2\right\rangle }$ ($V_{b}\simeq 0.9$ meV), the active competition between
the transport channels $\left\vert 1,-3/2\right\rangle ^{-}\overset{\uparrow
}{\longrightarrow }\left\vert 0,-2\right\rangle $ and $\left\vert
1,-5/2\right\rangle \overset{\downarrow }{\longrightarrow }\left\vert
0,-2\right\rangle $ is destroyed by the new transport channel $\left\vert
2,-2\right\rangle \overset{\uparrow }{\longrightarrow }\left\vert
1,-5/2\right\rangle $ due to the electron tunneling process via the
transport channel $\left\vert 1,-5/2\right\rangle $ $\overset{\uparrow }{%
\longrightarrow }\left\vert 2,-2\right\rangle \overset{\uparrow }{%
\longrightarrow }\left\vert 1,-5/2\right\rangle $ can occur, which is
responsible for the super-Poissonian shot noise being decreased to the
sub-Poissonian distribution.

As for the $V_{g}=0.6$ case, the transport channel $\left\vert
2,2\right\rangle \overset{\downarrow }{\longrightarrow }\left\vert
1,5/2\right\rangle $, which is a slow transport channel for the $\Gamma
_{R}^{\uparrow }\gg \Gamma _{R}^{\downarrow }$ case, first participates in
the electron transport, see Figs. 5(c) and 5(d). With the bias voltage increasing up
to about $0.4$ meV, the fast electron tunneling process via the transport channel $%
\left\vert 2,-2\right\rangle \overset{\uparrow }{\longrightarrow }\left\vert
1,-5/2\right\rangle $ takes place, thus, the effective competition between fast and slow
transport channels can form, and the shot noise is rapidly enhanced and
reaches up to a relatively large super-Poissonian value. However, the new
transport channels $\left\vert 2,-2\right\rangle \overset{\downarrow }{%
\longrightarrow }\left\vert 1,-3/2\right\rangle ^{-}$ and $\left\vert
2,-1\right\rangle \overset{\uparrow }{\longrightarrow }\left\vert
1,-3/2\right\rangle ^{-}$ can be quickly opened with the bias voltage
further increasing, then the fast transport channel $\left\vert
2,-2\right\rangle \overset{\uparrow }{\longrightarrow }\left\vert
1,-5/2\right\rangle $ will be weakened when a spin-down electron tunnels
out the SMM through the transition from the eigenstates $\left\vert
2,-2\right\rangle $ to $\left\vert 1,-3/2\right\rangle ^{-}$, so that the
formed effective competition between fast and slow transport channels is suppressed and even destroyed.
Moreover, when the transport channel $\left\vert 2,2\right\rangle \overset%
{\downarrow }{\longrightarrow }\left\vert 1,5/2\right\rangle $ does not
participate in the quantum transport originating from the occupation probabilities $%
P_{\left\vert 2,2\right\rangle }$ and $P_{\left\vert 1,5/2\right\rangle }$
being approaching zero, the two transport channels $\left\vert
2,-2\right\rangle \overset{\uparrow }{\longrightarrow }\left\vert
1,-5/2\right\rangle $ and $\left\vert 2,-2\right\rangle \overset{\downarrow }%
{\longrightarrow }\left\vert 1,-3/2\right\rangle ^{-}$ can not form a new
effective competition between fast and slow transport channels due to a relatively fast electron
tunneling process via $\left\vert 2,-2\right\rangle \overset{\downarrow }{%
\longrightarrow }\left\vert 1,-3/2\right\rangle ^{-}\overset{\uparrow }{%
\longrightarrow }$ $\left\vert 2,-1\right\rangle \overset{\uparrow }{%
\longrightarrow }\left\vert 1,-3/2\right\rangle ^{-}$ can take place. Consequently, the super-Poissonian shot noise is decreased quickly to a
sub-Poissonian value and displays a sharp peak.

\subsection*{The ferromagnetic lead (Source) - SMM - ferromagnetic lead (Drain)}

We now consider the F-SMM-F system, the strengths of the spin-dependent
SMM-electrode coupling are characterized by $\Gamma _{\alpha }^{\uparrow
}=\Gamma (1+p_{\alpha })/2$ and $\Gamma _{\alpha }^{\downarrow }=\Gamma
(1-p_{\alpha })/2$, here we set $p_{L}=p_{R}=p$. For a small enough or
negative gate voltage and relatively large spin polarization of the source
and drain electrodes $p$, an obvious NDC is observed but weaker than that
in the N-SMM-F system, especially for a large enough spin polarization $p$,
see the solid and dashed lines in Figs. 4(a) and 6(a), and 4(e) and 6(e).
While for a relatively large gate voltage, such as $V_{g}\geq 0.4$ meV, a
weak NDC can be observed for a large enough spin polarization $p$, but that
in the N-SMM-F system does not occur. Interestingly, for a small enough or
negative gate voltage, the shot noise in the NDC region is dramatically
enhanced and reaches up to a super-Poissonian value, see the solid and
dashed lines in Figs. 6(b) and 6(f); whereas for a large enough gate voltage
the formed super-Poissonian shot noise in the NDC region is decreased, see
the short dashed, short dash-dotted and thick dashed lines in Fig. 6(f).
This characteristic depends on the formation mechanism of the NDC,
which is illustrated by the examples of $V_{g}=-0.1$ and $V_{g}=0.6$ with $%
p=0.9$.

For a negative gate voltage $V_{g}=-0.1$, the fast transport channel $%
\left\vert 1,5/2\right\rangle \overset{\uparrow }{\longrightarrow }%
\left\vert 0,2\right\rangle $ first enters the bias voltage window. When the
bias voltage increases up to about $0.48$ meV, the new spin-up electron tunneling
processes, namely, $\left\vert 1,-3/2\right\rangle ^{-}\overset{\uparrow }{%
\longrightarrow }\left\vert 0,-2\right\rangle $, $\left\vert
1,-1/2\right\rangle ^{-}\overset{\uparrow }{\longrightarrow }\left\vert
0,-1\right\rangle $, $\left\vert 1,1/2\right\rangle ^{-}\overset{\uparrow }{%
\longrightarrow }\left\vert 0,0\right\rangle $, $\left\vert
1,3/2\right\rangle ^{-}\overset{\uparrow }{\longrightarrow }\left\vert
0,1\right\rangle $, and the spin-dowm electron tunneling processes, namely, $%
\left\vert 1,-5/2\right\rangle \overset{\downarrow }{\longrightarrow }%
\left\vert 0,-2\right\rangle $, $\left\vert 1,-3/2\right\rangle ^{-}\overset{%
\downarrow }{\longrightarrow }\left\vert 0,-1\right\rangle $, $\left\vert
1,-1/2\right\rangle ^{-}\overset{\downarrow }{\longrightarrow }\left\vert
0,0\right\rangle $, $\left\vert 1,1/2\right\rangle ^{-}\overset{\downarrow }{%
\longrightarrow }\left\vert 0,1\right\rangle $, $\left\vert
1,3/2\right\rangle ^{-}\overset{\downarrow }{\longrightarrow }\left\vert
0,2\right\rangle $ begin to participate in the quantum transport, see Figs.
7(a) and 7(b). This leads to the current magnitude of the fast transport channel $%
\left\vert 1,5/2\right\rangle \overset{\uparrow }{\longrightarrow }%
\left\vert 0,2\right\rangle $ decrease, but the increased current magnitudes
of the new opened transport channels are too small to compensate the
decreased current magnitude of $\left\vert 1,5/2\right\rangle \overset{%
\uparrow }{\longrightarrow }\left\vert 0,2\right\rangle $. Thus,
a NDC region can form, in which the corresponding shot noise is rapidly
enhanced by the active competition between the fast channel of current
decreasing and the slow channels of current increasing, and reaches up to a
large super-Poissonian value, see the solid line in Fig. 6(f). With further
increasing the bias voltage, the formed active competition between the fast
channel of current decreasing and the slow channels of current increasing
is weakened and even disappears, but the effective competition between the
spin-up and spin-down electron tunneling processes is still valid due to $%
\Gamma _{L}^{\uparrow }\gg \Gamma _{L}^{\downarrow }$ and $\Gamma
_{R}^{\uparrow }\gg \Gamma _{R}^{\downarrow }$, thus, the value of the
formed super-Poissonian begins to continually decrease but still remains
super-Poissonian distribution. When the bias voltage increases up to $0.8$
meV, the current magnitudes of the transport channels originating from the
transitions between the double- and singly-occupied eigenstates are already
larger than that of the some transport channels originating from the
transitions between the singly- and empty-occupied eigenstates, for example,
$\left\vert 2,2\right\rangle \overset{\uparrow }{\longrightarrow }\left\vert
1,3/2\right\rangle ^{+}$. In this case, the formed effective competition between the fast and slow
transport channels is suppressed and finally destroyed due to these
transport channels via the transitions from the double- to singly-occupied
eigenstates entering the bias voltage. Consequently, the super-Poissonian shot
noise is decreased quickly up to a sub-Poissonian value, see the solid line
in Fig. 6(f).

Compared with the $V_{g}=-0.1$ case, for $V_{g}=0.6$ the transport channel $%
\left\vert 2,2\right\rangle \overset{\downarrow }{\longrightarrow }%
\left\vert 1,5/2\right\rangle $ first participates in the quantum transport.
When the bias voltage increases up to about $4.8$ meV, the spin-up transport
channels $\left\vert 2,-2\right\rangle \overset{\uparrow }{\longrightarrow }%
\left\vert 1,-5/2\right\rangle $, $\left\vert 2,-1\right\rangle \overset{%
\uparrow }{\longrightarrow }\left\vert 1,-3/2\right\rangle ^{-}$, $%
\left\vert 2,0\right\rangle \overset{\uparrow }{\longrightarrow }\left\vert
1,-1/2\right\rangle ^{-}$, $\left\vert 2,1\right\rangle \overset{\uparrow }{%
\longrightarrow }\left\vert 1,1/2\right\rangle ^{-}$, $\left\vert
2,2\right\rangle \overset{\uparrow }{\longrightarrow }\left\vert
1,3/2\right\rangle ^{-}$, and the spin-down transport channels $\left\vert
2,-2\right\rangle \overset{\downarrow }{\longrightarrow }\left\vert
1,-3/2\right\rangle ^{-}$, $\left\vert 2,-1\right\rangle \overset{\downarrow
}{\longrightarrow }\left\vert 1,-1/2\right\rangle ^{-}$, $\left\vert
2,0\right\rangle \overset{\downarrow }{\longrightarrow }\left\vert
1,1/2\right\rangle ^{-}$, $\left\vert 2,1\right\rangle \overset{\downarrow }{%
\longrightarrow }\left\vert 1,3/2\right\rangle ^{-}$ can be opened, while
the current magnitude of the transport channel $\left\vert 2,2\right\rangle
\overset{\downarrow }{\longrightarrow }\left\vert 1,5/2\right\rangle $ begin
to decrease, see Figs. 7(c) and 7(d). For the $\Gamma _{R}^{\uparrow }\gg \Gamma
_{R}^{\downarrow }$ case, the decreased current magnitude of the spin-down
transport channel $\left\vert 2,2\right\rangle \overset{\downarrow }{%
\longrightarrow }\left\vert 1,5/2\right\rangle $ is smaller than the
increased current magnitudes of the new opened transport channels, thus, the NDC
does not appear. Whereas the active competition between the fast channel of
current decreasing and the slow channels of current increasing in a
relatively small bias voltage range can form but soon be destroyed, so that
the shot noise is significantly enhanced up to a very large super-Poissonian
value, then this value begins to decrease but still remains super-Poissonian
distribution due to the effective competition between the spin-up and
spin-down electron tunneling processes being still valid, see the thick dashed
line in Fig. 6(f). In particular, it is interesting note that\
the current magnitudes of the transport channels $\left\vert 2,2\right\rangle
\overset{\downarrow }{\longrightarrow }\left\vert 1,5/2\right\rangle $ and $%
\left\vert 2,2\right\rangle \overset{\uparrow }{\longrightarrow }\left\vert
1,3/2\right\rangle ^{-}$ increase with further increasing the bias voltage,
while the current magnitudes of the other transport channels $\left\vert
2,-2\right\rangle \overset{\uparrow }{\longrightarrow }\left\vert
1,-5/2\right\rangle $, $\left\vert 2,-1\right\rangle \overset{\uparrow }{%
\longrightarrow }\left\vert 1,-3/2\right\rangle ^{-}$, $\left\vert
2,0\right\rangle \overset{\uparrow }{\longrightarrow }\left\vert
1,-1/2\right\rangle ^{-}$, $\left\vert 2,1\right\rangle \overset{\uparrow }{%
\longrightarrow }\left\vert 1,1/2\right\rangle ^{-}$, $\left\vert
2,-2\right\rangle \overset{\downarrow }{\longrightarrow }\left\vert
1,-3/2\right\rangle ^{-}$, $\left\vert 2,-1\right\rangle \overset{\downarrow
}{\longrightarrow }\left\vert 1,-1/2\right\rangle ^{-}$, $\left\vert
2,0\right\rangle \overset{\downarrow }{\longrightarrow }\left\vert
1,1/2\right\rangle ^{-}$ and $\left\vert 2,1\right\rangle \overset{%
\downarrow }{\longrightarrow }\left\vert 1,3/2\right\rangle ^{-}$ decrease.
This feature leads to the occurrence of a weak NDC. In this NDC bias voltage
range, however, the super-Poissonian shot noise value continually decreases,
see the thick dashed line in Fig. 6(f). When the transport channels
originating from the transitions from the singly- to empty-occupied
eigenstates enter the bias voltage, the physical mechanism of decreasing
super-Poissonian shot noise is the same as the $V_{g}=-0.1$ case, namely,
the formed effective competition between the spin-up and spin-down electron
tunneling processes is weakened even destroyed by these current increased
transport channels. This is responsible for the super-Poissonian shot noise
being decreased to a sub-Poissonian value.

We now study the skewness and kurtosis properties of the transport current in the
super-Poissonian shot noise bias voltage regions. It is well known that the skewness and
kurtosis (both its magnitude and sign) characterize, respectively, the asymmetry of and
the peakedness of the probability distribution around the average transferred-electron
number $\bar{n}$ during a time interval t, thus that provide further information for
the counting statistics beyond the shot noise. In the N-SMM-F system with a given small
enough or negative gate voltage, for a relatively large $p_{R}$, the skewness shows a crossover
from a large negative to a relatively small positive values, while the kurtosis shows a crossover
from a large positive to a relatively small negative values, see the solid and dashed lines in
Figs. 4(c) and 4(d); whereas for a large enough $p_{R}$, the transition of the skewness from
a large negative to a large positive values takes place and forms a Fano-like resonance,
see the solid, dashed and dotted lines in Fig. 4(g), while the transitions of the kurtosis from a large
positive to a large negative values and then from the large negative to a large positive values
take place, and form the double Fano-like resonances, see the solid, dashed and dotted lines in Fig. 4(h).
In contrast with a small enough or negative gate voltage, for a large enough gate voltage, the skewness and kurtosis for
a relatively large $p_{R}$ show, respectively, the crossovers from a large positive to
a relatively small negative values and from a small positive to a relatively large negative values, see the short
dash-dotted and thick dashed lines in Figs. 4(c) and 4(d); while for
a large enough $p_{R}$, the skewness and kurtosis show, respectively, the crossovers from a small positive to
a relatively large negative values and from a small negative large to a relatively large positive values, see the short
dash-dotted and thick dashed lines in Figs. 4(g) and 4(h), but the variations in the magnitudes of the
skewness and kurtosis are much smaller than that for a small enough or negative gate voltage, see Figs. 4(g) and 4(h).
As for the F-SMM-F system with a given relatively large $p$, the skewness for a small enough or negative gate voltage shows a large negative value, see the solid,
dashed and dotted lines in Figs. 6(c) and 6(g), whereas for a large enough gate voltage that shows
a large positive value, see the short dashed, short dash-dotted and thick dashed lines in Figs. 6(c) and 6(g);
while the kurtosis shows the double-crossover from a large positive to a relatively small negative
values and then from the negative to a large positive values but the latter has a remarkable
variation in the magnitude of the kurtosis, see Figs. 6(d) and 6(h).
Moreover, we found that the magnitudes of the skewness and kurtosis are more sensitive to the
active competition between the fast channels of current decreasing and the corresponding slow
channels of current increasing than the shot noise, see the short dashed, short dash-dotted and thick dashed lines in Figs. 2, 4 and 6.

\section*{Discussion}
In summary, we have studied the NDC and super-Poissonian shot noise
properties of electron transport through a SMM weakly coupled to two
electrodes with either one or both of them being ferromagnetic, and analyzed
the skewness and kurtosis properties of the transport current in the super-Poissonian
shot noise regions. It is demonstrated that the occurrences of the NDC and super-Poissonian shot noise
depend sensitively on the spin polarization of the soure and drain
electrodes and the applied gate voltage. For the F-SMM-N system, when the
transition from the double- to singly-occupied eigenstates first enters the
bias voltage window, which corresponds to a large enough gate voltage, the
super-Poissonian shot noise is observed for a large enough spin polarization
of left electrode $p_{L}$. As for the N-SMM-F system, the NDC and super-Poissonian shot
noise can be observed for a relatively large spin polarization of right
electrode $p_{R}$ and a small enough or negative gate volatge, especially
for a large enough $p_{R}$ a strong NDC and a very large value of the
super-Poissonian shot noise appear, and the shot noise in the NDC region is
first enhanced up to a super-Poissonian value and then is decreased but
still remains super-Poissonian distribution for a large enough $p_{R}$; while
for a large enough gate voltage and a relatively large $p_{R}$ the
super-Poissonian shot noise is only observed. Compared with the N-SMM-F
system, for the F-SMM-F system a relatively weak NDC and a large
super-Poissonian shot noise bias voltage range are observed; whereas the
formed super-Poissonian shot noise in the NDC region is continually
decreased for a large enough gate voltage and spin polarization of left and
right electrodes $p$. Furthermore, the transitions of the skewness and kurtosis
from a large positive (negative) to a large negative (positive) values are also observed,
which can provide a deeper and better understanding of electron
transport through single-molecule magnet junctions.
The observed NDC and super-Poissonian shot noise in the SMM system can be
qualitatively attributed to the effective competition between the fast and slow transport
channels, and the NDC properties suggest a gate-voltage-controlled NDC molecular device.

\section*{Methods}
The SMM-electrode coupling is assumed to be sufficiently weak, so that the
sequential tunneling is dominant. The transitions are well described by the
quantum master equation of a reduced density matrix spanned by the
eigenstates of the SMM. Under the second order Born approximation and Markov
approximation, the particle-number-resolved quantum master equation for the
reduced density matrix is given by \cite{Li1,Li2,WangSK}
\begin{equation}
\dot{\rho}^{\left( n\right) }\left( t\right) =-i\mathcal{L}\rho ^{\left(
n\right) }\left( t\right) -\frac{1}{2}\mathcal{R}\rho ^{\left( n\right)
}\left( t\right) ,  \label{Master1}
\end{equation}%
with%
\begin{align}
\mathcal{R}\rho ^{\left( n\right) }\left( t\right) & =%
{\displaystyle\sum\limits_{\mu=\uparrow,\downarrow}}
\left[ d_{\mu }^{\dagger }A_{\mu }^{\left( -\right)
}\rho ^{\left( n\right) }\left( t\right) +\rho ^{\left( n\right) }\left(
t\right) A_{\mu }^{\left( +\right) }d_{\mu }^{\dagger }-A_{L\mu }^{\left(
-\right) }\rho ^{\left( n\right) }\left( t\right) d_{\mu }^{\dagger }\right.
\notag \\
& \left. -d_{\mu }^{\dagger }\rho ^{\left( n\right) }\left( t\right) A_{L\mu
}^{\left( +\right) }-A_{R\mu }^{\left( -\right) }\rho ^{\left( n-1\right)
}\left( t\right) d_{\mu }^{\dagger }-d_{\mu }^{\dagger }\rho ^{\left(
n+1\right) }\left( t\right) A_{R\mu }^{\left( +\right) }\right] +H.c.,
\label{Master2}
\end{align}%
where $A_{\mu }^{\left( \pm \right) }=\sum_{\alpha =L,R}A_{\alpha \mu
}^{\left( \pm \right) }$, $A_{\alpha \mu }^{\left( \pm \right) }=\Gamma
_{\alpha }^{\mu }n_{\alpha }^{\left( \pm \right) }\left( -\mathcal{L}\right)
d_{\mu }$, $n_{\alpha }^{\left( +\right) }=f_{\alpha }$ and $n_{\alpha }^{\left(
-\right) }=1-f_{\alpha }$ ($f_{\alpha }$ is the Fermi function of the
electrode $\alpha $). Liouvillian superoperator $\mathcal{L}$ is defined as $%
\mathcal{L}\left( \cdots \right) =\left[ H_{\text{SMM}},\left( \cdots
\right) \right] $. $\rho ^{\left( n\right) }\left( t\right) $ is the reduced
density matrix of the SMM conditioned by the electron numbers arriving at
the right electrode up to time $t$. In order to calculate the first four
cumulants, one can define $S\left( \chi ,t\right) =\sum_{n}\rho ^{\left(
n\right) }\left( t\right) e^{in\chi }$. According to the definition of the
cumulant generating function $e^{-F\left( \chi \right) }=\sum_{n}$Tr$\left[
\rho ^{\left( n\right) }\left( t\right) \right] e^{in\chi }=\sum_{n}P\left(
n,t\right) e^{in\chi }$, we evidently have $e^{-F\left( \chi \right) }=$Tr$%
\left[ S\left( \chi ,t\right) \right] $, where the trace is over the
eigenstates of the SMM. Since Eq. (\ref{Master1}) has the following form
\begin{equation}
\dot{\rho}^{\left( n\right) }=A\rho ^{\left( n\right) }+C\rho ^{\left(
n+1\right) }+D\rho ^{\left( n-1\right) },  \label{formalmaster}
\end{equation}%
$S\left( \chi ,t\right) $ satisfies
\begin{equation}
\dot{S}=AS+e^{-i\chi }CS+e^{i\chi }DS\equiv \mathcal{L}_{\chi }S.
\label{formalmaster1}
\end{equation}%
In the low frequency limit, the counting time
is much longer than the time of electron tunneling through the SMM. In this
case, $F\left( \chi \right) $ can be expressed as\cite%
{Bagrets,Groth,Flindt01,Kieblich}
\begin{equation}
F\left( \chi \right) =-\lambda _{1}\left( \chi \right) t,  \label{CGFformal}
\end{equation}%
where $\lambda _{1}\left( \chi \right) $ is the eigenvalue of $\mathcal{L}%
_{\chi }$ which goes to zero for $\chi \rightarrow 0$. According to the
definition of the cumulants, one can express $\lambda _{1}\left( \chi \right)
$\ as
\begin{equation}
\lambda _{1}\left( \chi \right) =\frac{1}{t}\sum_{k=1}^{\infty }C_{k}\frac{%
\left( i\chi \right) ^{k}}{k!}.  \label{Lambda}
\end{equation}%
Here, the first four cumulants $C_{k}$ are directly related to
the transport characteristics. For example, the first-order cumulant (the
peak position of the distribution of transferred-electron number) $C_{1}=%
\bar{n}$ gives the average current $\left\langle I\right\rangle =eC_{1}/t$.
The zero-frequency shot noise is related to the second-order cumulant (the
peak-width of the distribution) $S=2e^{2}C_{2}/t=2e^{2}\left( \overline{n^{2}%
}-\bar{n}^{2}\right) /t$. The third-order $C_{3}=\overline{\left( n-\bar{n}%
\right) ^{3}}$ and four-order $C_{4}=\overline{\left( n-\bar{n}\right) ^{4}}$%
characterize, respectively, the skewness and kurtosis of the distribution.
Here, $\overline{\left( \cdots \right) }=\sum_{n}\left( \cdots \right)
P\left( n,t\right) $. In general, the shot noise, skewness and kurtosis are
represented by the Fano factor $F_{2}=C_{2}/C_{1}$, $F_{3}=C_{3}/C_{1}$ and $%
F_{4}=C_{4}/C_{1}$, respectively.

The low order cumulants can be calculated by the Rayleigh--Schr%
\"{o}dinger perturbation theory in the counting parameter $\chi $. In order
to calculate the first four current cumulants we expand $L_{\chi }$ to four
order in $\chi $%
\begin{equation}
L_{\chi }=L_{0}+L_{1}\chi +\frac{1}{2!}L_{2}\chi ^{2}+\frac{1}{3!}L_{3}\chi
^{3}+\frac{1}{4!}L_{4}\chi ^{4}\cdots .  \label{matirxL}
\end{equation}%
and define the two projectors\cite{Flindt01,Flindt02,Flindt03} $P=P^{2}=\left\vert \left. 0\right\rangle
\right\rangle \left\langle \left\langle \tilde{0}\right. \right\vert $ and $%
Q=Q^{2}=1-P$, obeying the relations $PL_{0}=L_{0}P=0$ and $%
QL_{0}=L_{0}Q=L_{0}$. Here, $\left\vert \left. 0\right\rangle \right\rangle $
being the steady state $\rho ^{stat}$ is the right eigenvector of $L_{0}$,
namely, $L_{0}\left\vert \left. 0\right\rangle \right\rangle =0$, and $%
\left\langle \left\langle \tilde{0}\right. \right\vert \equiv \hat{1}$ is
the corresponding left eigenvector. In view of $L_{0}$ being regular, we also
introduce the pseudoinverse according to $R=QL_{0}^{-1}Q$, which is
well-defined due to the inversion being performed only in the subspace spanned
by $Q$. After a careful calculation, $\lambda _{1}\left( \chi \right) $ is
given by
\begin{align}
\lambda _{1}\left( \chi \right) & =\left\langle \left\langle \tilde{0}%
\right. \right\vert L_{1}\left\vert \left. 0\right\rangle \right\rangle \chi
\notag \\
& +\frac{1}{2!}\left[ \left\langle \left\langle \tilde{0}\right. \right\vert
L_{2}\left\vert \left. 0\right\rangle \right\rangle -2\left\langle
\left\langle \tilde{0}\right. \right\vert L_{1}RL_{1}\left\vert \left.
0\right\rangle \right\rangle \right] \chi ^{2}  \notag \\
& +\frac{1}{3!}\left[ \left\langle \left\langle \tilde{0}\right. \right\vert
L_{3}\left\vert \left. 0\right\rangle \right\rangle -3\left\langle
\left\langle \tilde{0}\right. \right\vert \left(
L_{2}RL_{1}+L_{1}RL_{2}\right) \left\vert \left. 0\right\rangle
\right\rangle \right.  \notag \\
& \left. -6\left\langle \left\langle \tilde{0}\right. \right\vert
L_{1}R\left( RL_{1}P-L_{1}R\right) L_{1}\left\vert \left. 0\right\rangle
\right\rangle \right] \chi ^{3}+  \notag \\
& +\frac{1}{4!}\left[ \left\langle \left\langle \tilde{0}\right. \right\vert
L_{4}\left\vert \left. 0\right\rangle \right\rangle -6\left\langle
\left\langle \tilde{0}\right. \right\vert L_{2}RL_{2}\left\vert \left.
0\right\rangle \right\rangle \right.  \notag \\
& -4\left\langle \left\langle \tilde{0}\right. \right\vert \left(
L_{3}RL_{1}+L_{1}RL_{3}\right) \left\vert \left. 0\right\rangle \right\rangle
\notag \\
& -12\left\langle \left\langle \tilde{0}\right. \right\vert L_{2}R\left(
RL_{1}P-L_{1}R\right) L_{1}\left\vert \left. 0\right\rangle \right\rangle
\notag \\
& -12\left\langle \left\langle \tilde{0}\right. \right\vert L_{1}R\left(
RL_{2}P-L_{2}R\right) L_{1}\left\vert \left. 0\right\rangle \right\rangle
\notag \\
& -12\left\langle \left\langle \tilde{0}\right. \right\vert L_{1}R\left(
RL_{1}P-L_{1}R\right) L_{2}\left\vert \left. 0\right\rangle \right\rangle
\notag \\
& -24\left\langle \left\langle \tilde{0}\right. \right\vert L_{1}R\left(
R^{2}L_{1}PL_{1}P-RL_{1}PL_{1}R-L_{1}R^{2}L_{1}P\right.  \notag \\
& \left. \left. -RL_{1}RL_{1}P+L_{1}RL_{1}R\right) L_{1}\left\vert \left.
0\right\rangle \right\rangle \right] \chi ^{4}+\cdots .  \label{matrixLambda}
\end{align}%
From Eqs. (\ref{Lambda}) and (\ref{matrixLambda}) we can identify the first
four current cumulants:%
\begin{equation}
C_{1}/t=\left\langle \left\langle \tilde{0}\right. \right\vert
L_{1}\left\vert \left. 0\right\rangle \right\rangle /i,  \label{current}
\end{equation}%
\begin{equation}
C_{2}/t=\left[ \left\langle \left\langle \tilde{0}\right. \right\vert
L_{2}\left\vert \left. 0\right\rangle \right\rangle -2\left\langle
\left\langle \tilde{0}\right. \right\vert L_{1}RL_{1}\left\vert \left.
0\right\rangle \right\rangle \right] /i^{2},  \label{shot noise}
\end{equation}%
\begin{align}
& C_{3}/t=\left[ \left\langle \left\langle \tilde{0}\right. \right\vert
L_{3}\left\vert \left. 0\right\rangle \right\rangle -3\left\langle
\left\langle \tilde{0}\right. \right\vert \left(
L_{2}RL_{1}+L_{1}RL_{2}\right) \left\vert \left. 0\right\rangle
\right\rangle \right.  \notag \\
& \left. -6\left\langle \left\langle \tilde{0}\right. \right\vert
L_{1}R\left( RL_{1}P-L_{1}R\right) L_{1}\left\vert \left. 0\right\rangle
\right\rangle \right] /i^{3}.  \label{skewness}
\end{align}%
\begin{align}
& C_{4}/t=\left[ \left\langle \left\langle \tilde{0}\right. \right\vert
L_{4}\left\vert \left. 0\right\rangle \right\rangle -6\left\langle
\left\langle \tilde{0}\right. \right\vert L_{2}RL_{2}\left\vert \left.
0\right\rangle \right\rangle \right.  \notag \\
& -4\left\langle \left\langle \tilde{0}\right. \right\vert \left(
L_{3}RL_{1}+L_{1}RL_{3}\right) \left\vert \left. 0\right\rangle \right\rangle
\notag \\
& -12\left\langle \left\langle \tilde{0}\right. \right\vert L_{2}R\left(
RL_{1}P-L_{1}R\right) L_{1}\left\vert \left. 0\right\rangle \right\rangle
\notag \\
& -12\left\langle \left\langle \tilde{0}\right. \right\vert L_{1}R\left(
RL_{2}P-L_{2}R\right) L_{1}\left\vert \left. 0\right\rangle \right\rangle
\notag \\
& -12\left\langle \left\langle \tilde{0}\right. \right\vert L_{1}R\left(
RL_{1}P-L_{1}R\right) L_{2}\left\vert \left. 0\right\rangle \right\rangle
\notag \\
& -24\left\langle \left\langle \tilde{0}\right. \right\vert L_{1}R\left(
R^{2}L_{1}PL_{1}P-RL_{1}PL_{1}R-L_{1}R^{2}L_{1}P\right.  \notag \\
& \left. \left. -RL_{1}RL_{1}P+L_{1}RL_{1}R\right) L_{1}\left\vert \left.
0\right\rangle \right\rangle \right] /i^{4}.  \label{kurtosis}
\end{align}%
The above four equations are the starting point of the numerical calculation.

\section*{Acknowledgments}
This work was supported by the NKBRSFC under grants Nos. 2011CB921502, 2012CB821305,
NSFC under grants Nos. 11204203, 61405138, 11275118, 61227902, 61378017, 11434015,
SKLQOQOD under grants No. KF201403, SPRPCAS under grants No. XDB01020300.

\section*{Author Contributions}
H. B. X. conceived the idea and designed the research and performed calculations.
J. Q. L. and W. M. L. contributed to the analysis and interpretation of the results and prepared the manuscript.

\section*{Competing Interests}
The authors declare no competing financial interests.

\section*{Correspondence}
Correspondence and requests for materials should be addressed to Hai-Bin Xue or Wu-Ming Liu.

\clearpage
\newpage

\begin{figure*}[t]
\centerline{\includegraphics[height=12cm,width=16cm]{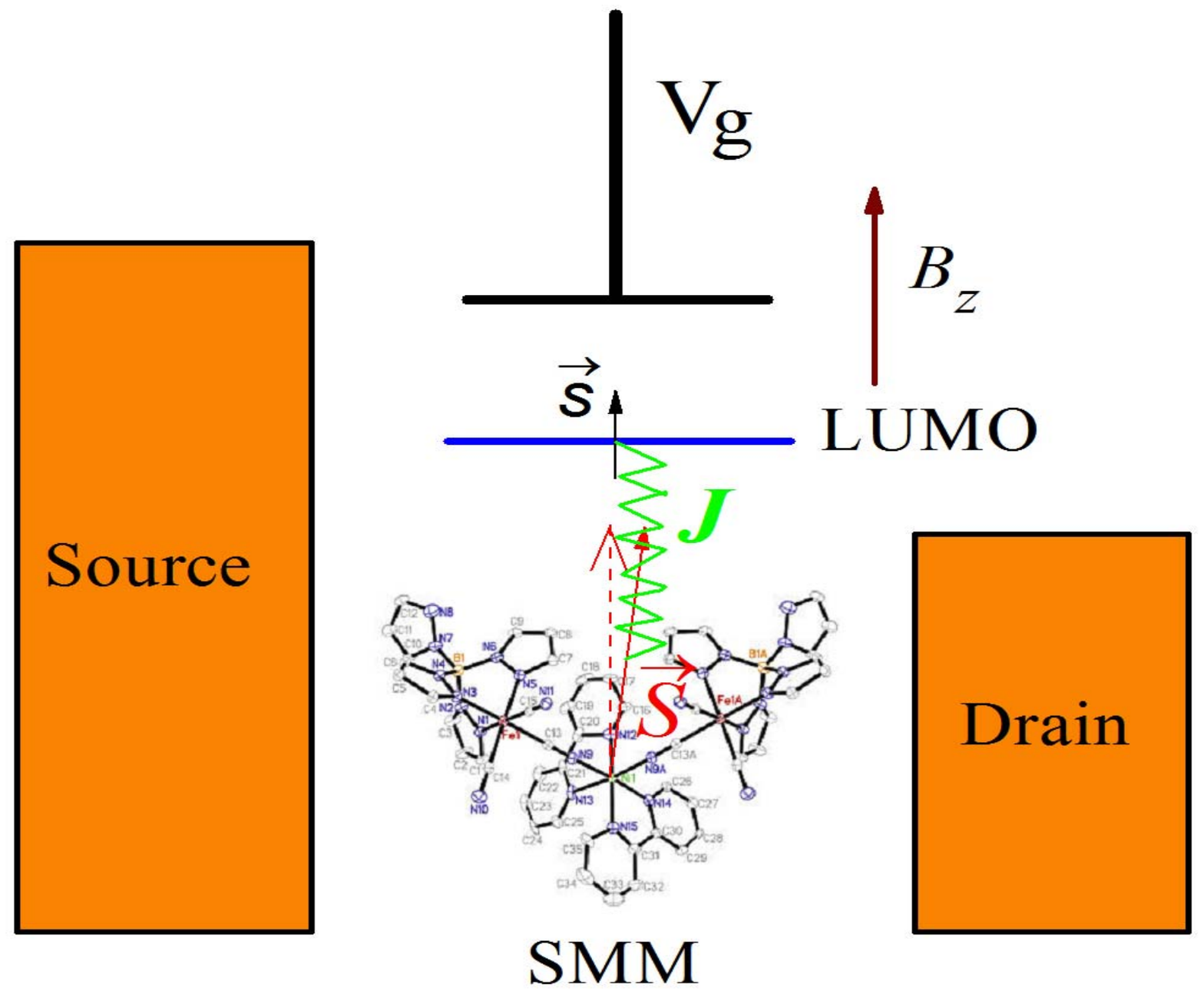}}
\caption{Schematic representation of a single-molecule magnet (SMM) weakly coupled to two leads.
The SMM consists of the lowest unoccupied non-degenerate molecular orbital (LUMO),
which can be tuned by a gate voltage $V_{g}$, the phenomenological giant
spin $\protect\overrightarrow{S}$, and the uniaxial anisotropy energy $K_{2}(S_{z})^{2}$. The exchange coupling
between the conduction electron spin $\vec{s}$ in the LUMO and the SMM spin $\vec{S}$ is
denoted by $J$. The external magnetic field $B_{z}$ is applied along the easy axis of the SMM.
Here, we consider three different electrode configurations,namely, (i) the ferromagnetic
lead (Source) - SMM - normal-metal lead (Drain), (ii) the normal-metal
lead (Source) - SMM - ferromagnetic lead (Drain), (iii) the ferromagnetic
lead (Source) - SMM - ferromagnetic lead (Drain). } \label{fig1}
\end{figure*}

\begin{figure*}[t]
\centerline{\includegraphics[height=14cm,width=16cm]{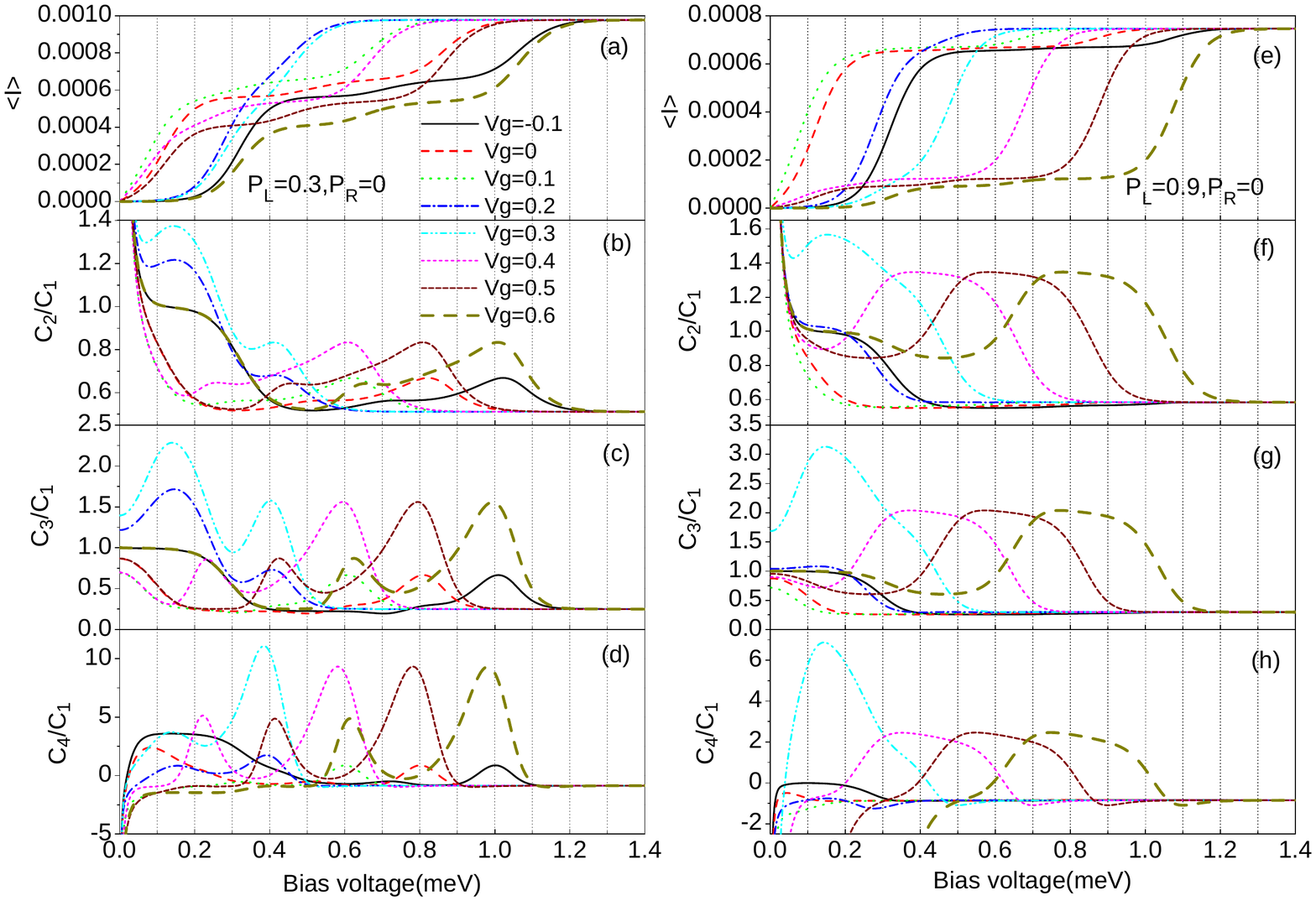}}
  \caption{The average current ($\left\langle I\right\rangle$), shot noise ($C_{2}/C_{1}$),
skewness ($C_{3}/C_{1}$) and kurtosis ($C_{4}/C_{1}$) vs bias voltage for
different gate voltages, here $C_{k}$ is the zero-frequency ${k}$-order cumulant of current fluctuations. (a), (b), (c) and (d) for $p_{L}=0.3$ and $p_{R}=0$;
(e), (f), (g) and (h) for $p_{L}=0.9$, $p_{R}=0$. The SMM parameters: $S=2$, $\varepsilon
_{d}=0.2$, $U=0.1$, $J=0.1$, $K_{2}=0.04$, $B=0.08$, $\Gamma _{L}=\Gamma
_{R}=\Gamma=0.002$ and $k_{B}T=0.02$.} \label{fig2}
\end{figure*}

\begin{figure*}[t]
\centerline{\includegraphics[height=12cm,width=16cm]{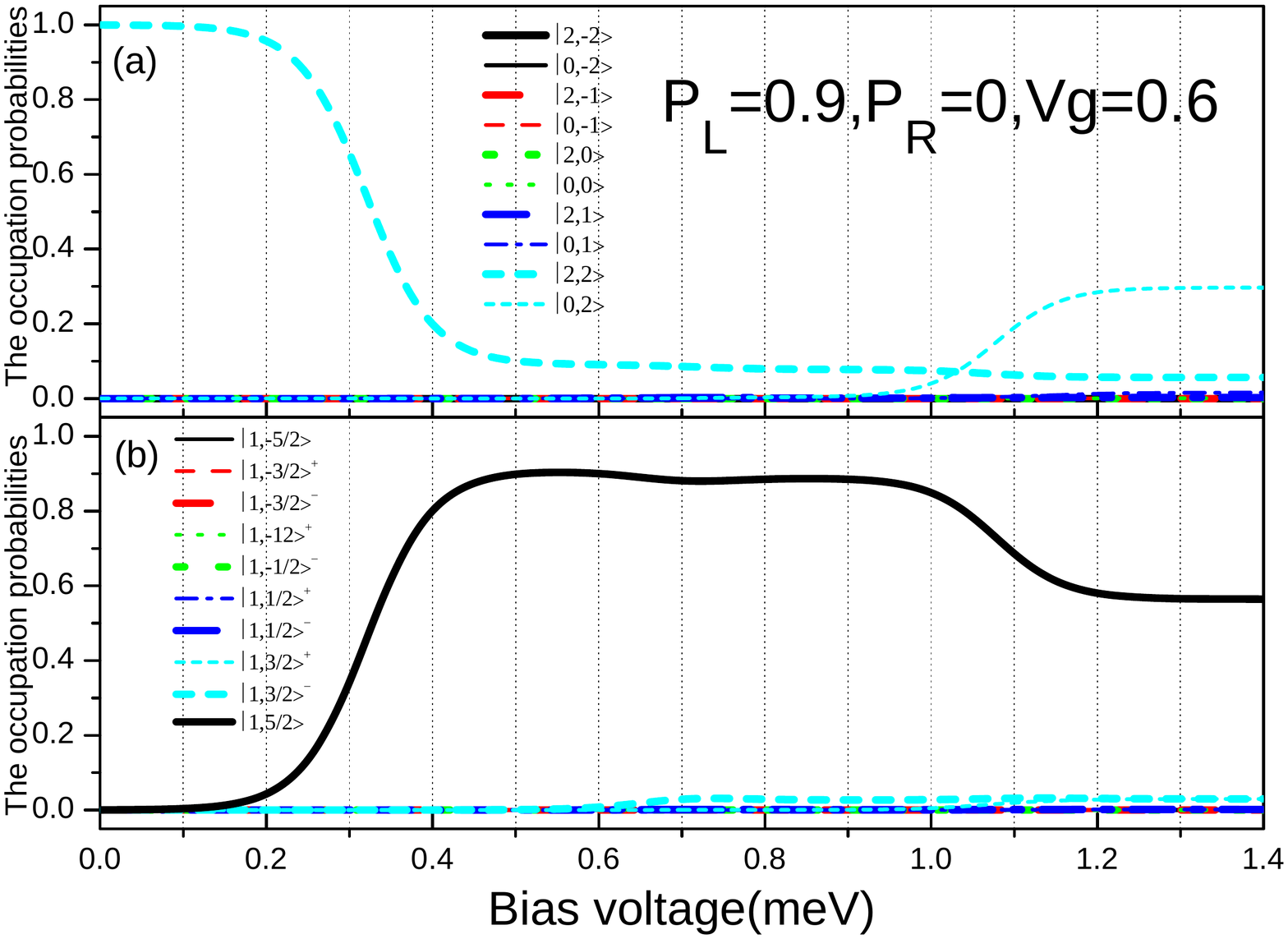}}
  \caption{The occupation probabilities of the SMM eigenstates vs
bias voltage for $p_{L}=0.9$, $p_{R}=0$ and $V_{g}=0.6$. The SMM parameters
are the same as in Fig. 1.} \label{fig3}
\end{figure*}

\begin{figure*}[t]
\centerline{\includegraphics[height=14cm,width=16cm]{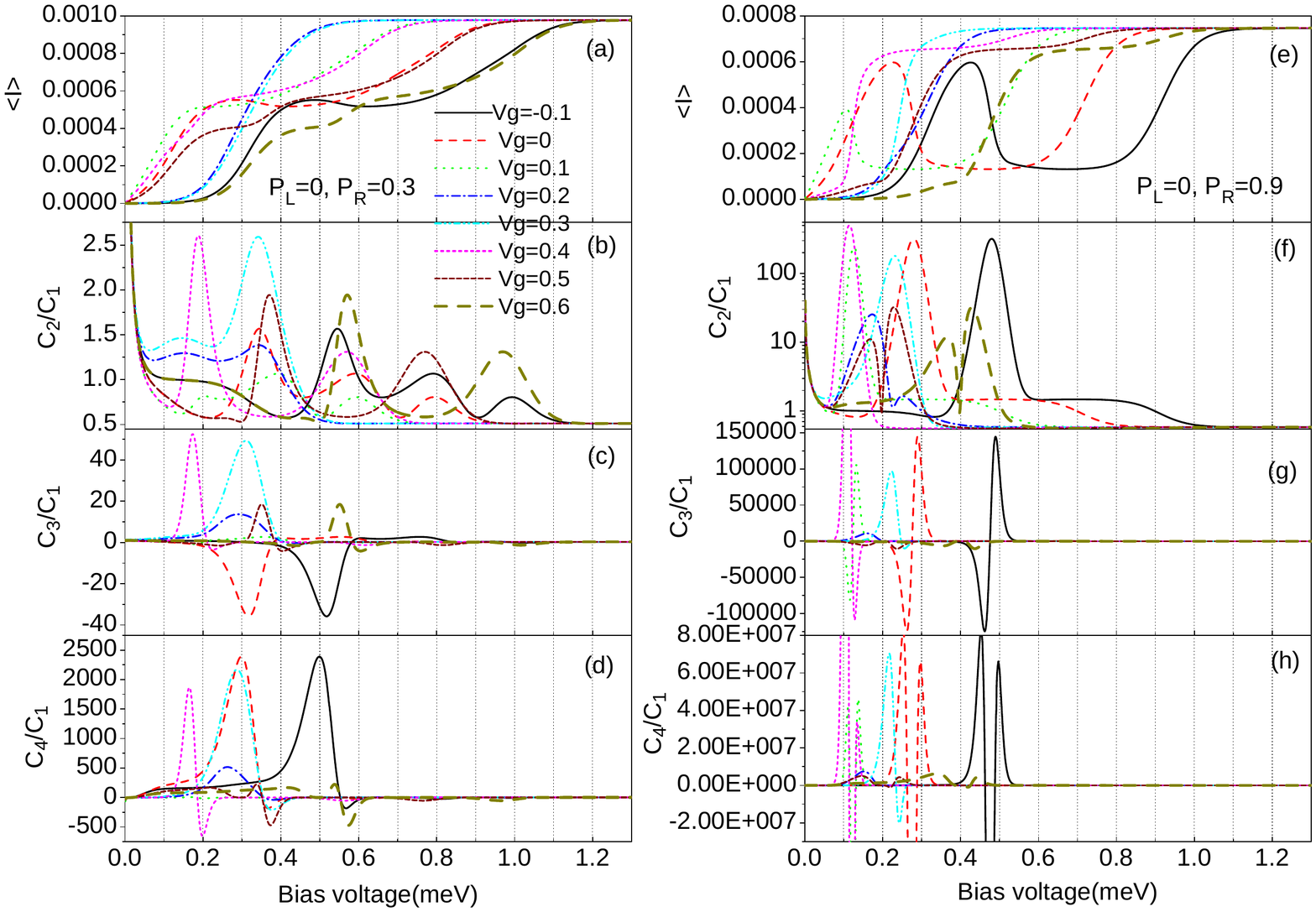}}
  \caption{The average current ($\left\langle I\right\rangle$), shot noise ($C_{2}/C_{1}$), skewness ($C_{3}/C_{1}$) and
kurtosis ($C_{4}/C_{1}$) vs bias voltage for
different gate voltages, here $C_{k}$ is the zero-frequency ${k}$-order cumulant of current fluctuations. (a), (b), (c) and (d) for $p_{L}=0$ and $p_{R}=0.3$;
(e), (f), (g) and (h) for $p_{L}=0$, $p_{R}=0.9$. The SMM parameters are the same as in Fig. 1.} \label{fig4}
\end{figure*}

\begin{figure*}[t]
\centerline{\includegraphics[height=14cm,width=16cm]{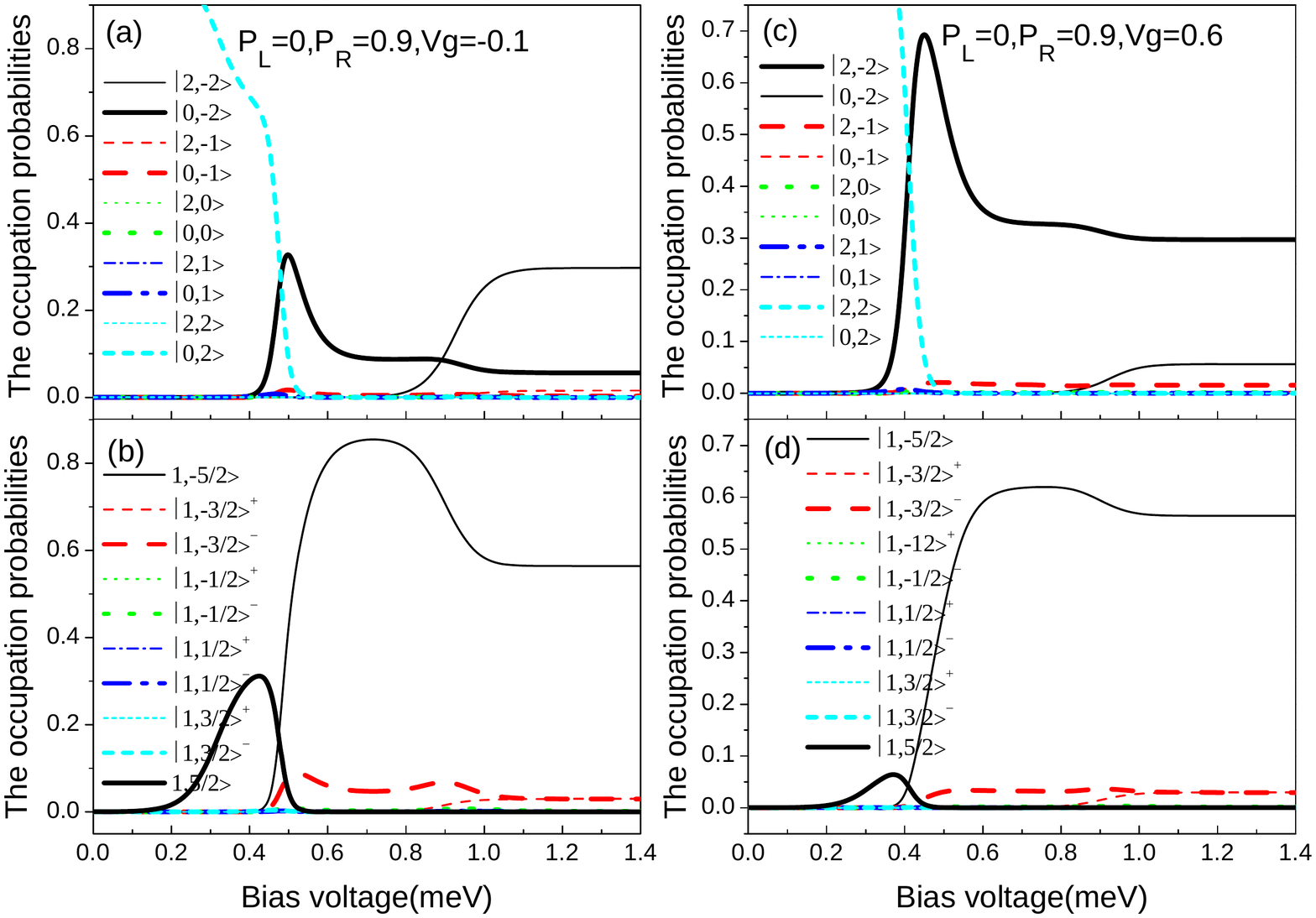}}
  \caption{The occupation probabilities of the SMM eigenstates vs
bias voltage for different gate voltages with $p_{L}=0$ and $p_{R}=0.9$. (a)
and (b) for $V_{g}=-0.1$; (c) and (d) for $V_{g}=0.6$. The SMM parameters
are the same as in Fig. 1.} \label{fig5}
\end{figure*}

\begin{figure*}[t]
\centerline{\includegraphics[height=14cm,width=16cm]{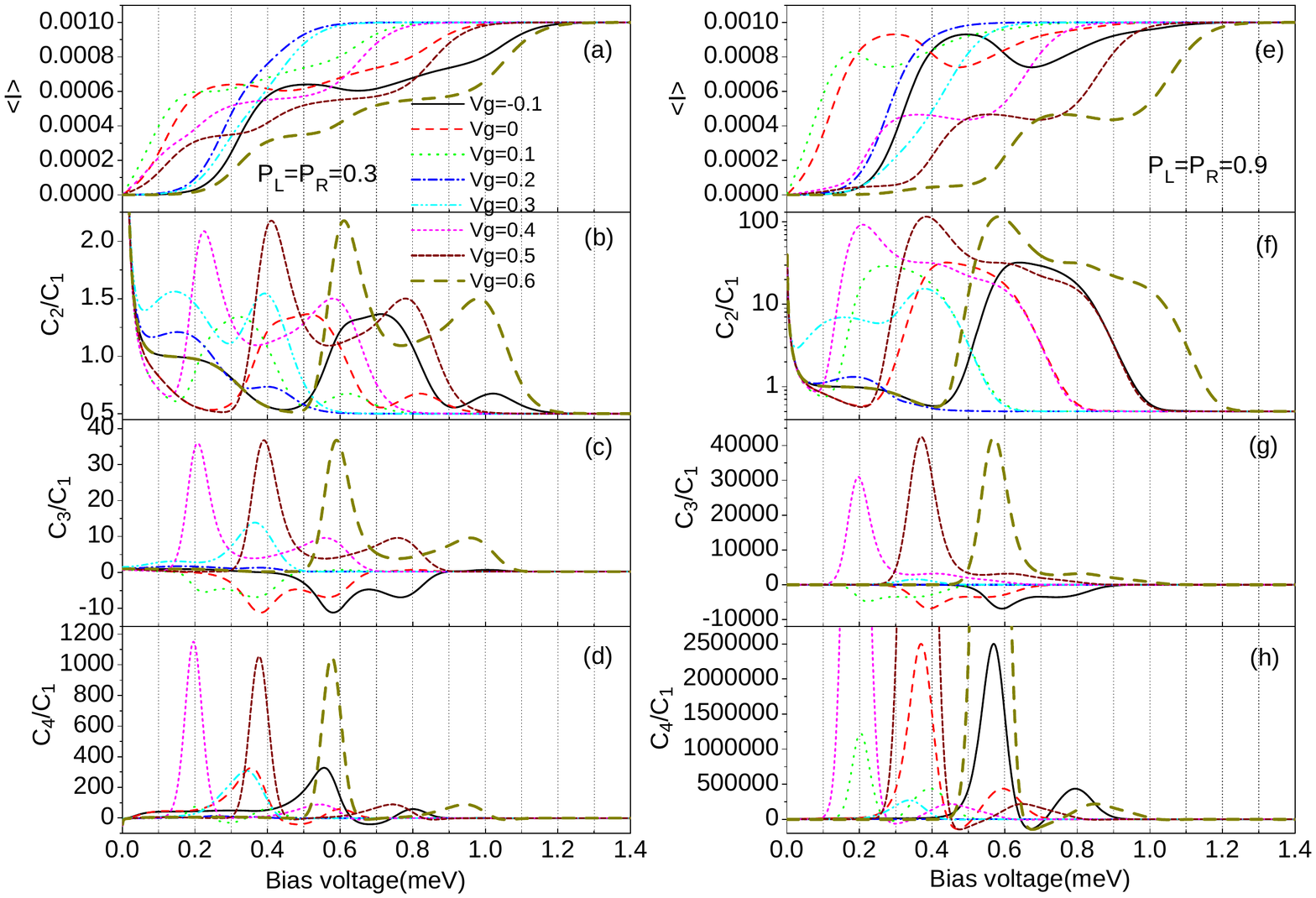}}
  \caption{The average current ($\left\langle I\right\rangle$), shot noise ($C_{2}/C_{1}$),
skewness ($C_{3}/C_{1}$) and kurtosis ($C_{4}/C_{1}$) vs bias voltage for
different gate voltages, here $C_{k}$ is the zero-frequency ${k}$-order cumulant of current fluctuations. (a), (b), (c) and (d) for $p_{L}=p_{R}=0.3$; (e), (f), (g) and (h) for $%
p_{L}=p_{R}=0.9$. The SMM parameters are the same as in Fig. 1.} \label{fig6}
\end{figure*}

\begin{figure*}[t]
\centerline{\includegraphics[height=14cm,width=16cm]{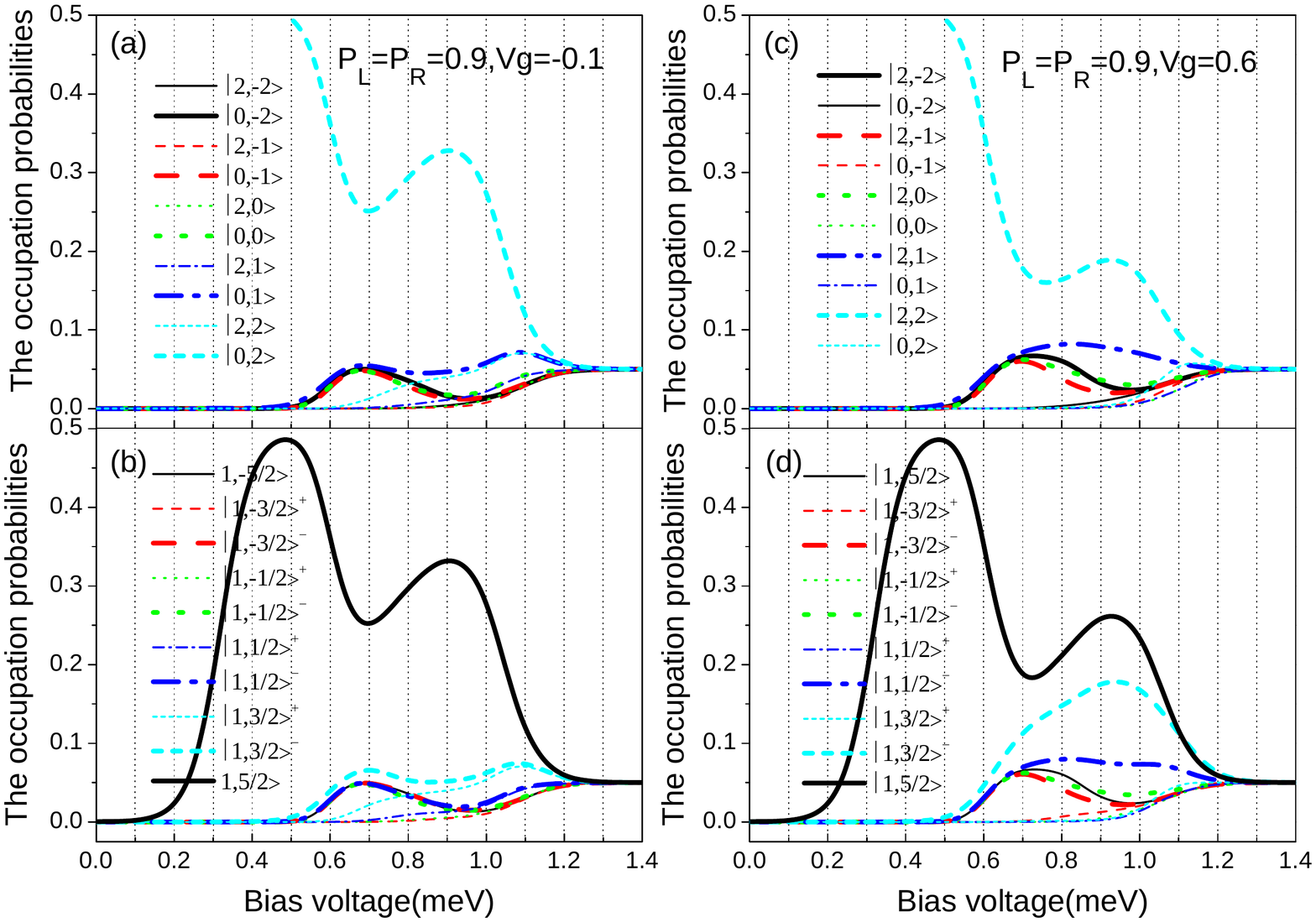}}
  \caption{The occupation probabilities of the SMM eigenstates vs
bias voltage for different gate voltages with $p_{L}=p_{R}=0.9$. (a) and (b)
for $V_{g}=-0.1$; (c) and (d) for $V_{g}=0.6$. The SMM parameters are the
same as in Fig. 1.} \label{fig7}
\end{figure*}

\end{document}